\documentclass[twocolumn,showpacs,amsmath,amssymb,floatfix,pre]{revtex4}
\usepackage{graphicx}

\begin{document}
\draft   

\title{Cooling rate, heating rate and aging effects in glassy water}
\author{Nicolas Giovambattista$^1$,
 H. Eugene Stanley$^1$ and Francesco Sciortino$^2$}
\address{$^1$Center for Polymer Studies and Department of Physics\\
  Boston University, Boston, MA 02215 USA\\
$^2$ Dipartimento di Fisica, Istituto Nazionale per la Fisica della
Materia,\\
and I.N.F.M. Center for Statistical Mechanics and Complexity,\\
Universit\`{a} di Roma La Sapienza, P.le A. Moro 2, I-00185 Roma, ITALY}


\begin{abstract}
We report a molecular dynamics simulation study of the properties
 of the potential energy landscape
sampled by a system of water molecules during the process of generating a
glass by  cooling, and during the process of regenerating the equilibrium
liquid by heating the glass. We study the dependence of these processes on
the cooling/heating rates as well as on the role of aging (the time elapsed in the glass state). We compare the properties of the potential energy landscape sampled during these processes with the corresponding properties sampled in the liquid equilibrium state to elucidate under which conditions  glass configurations can be associated with equilibrium liquid configurations. 
\end{abstract}

\maketitle


One recent activity in the physics of supercooled liquids and
glasses\cite{pablo,confbook} is the search for the conditions
 under which  a
glass can be considered a liquid whose structural properties have been
``frozen'' during the preparation process. If the glass can be connected to a
liquid state, then a thermodynamic  description of the glass state can be
developed\cite{teo,parisimezard,speedy,mossa}. 
Many routes can bring a system to an arrested disordered state
\cite{angellscience95}, such as vapor deposition, pressure induced
 amorphization, hyperquenching or standard cooling. Only the last two provide
 a continuous path from the liquid to the glass state and hence are
the best candidates for studying the connection between glass and liquid 
configurations.

Since water can be glassified by cooling only using
hyperquenching techniques (i.e. with rates of the order of
 $10^5$~K/s\cite{naturejohari}), understanding the connection between
 the liquid state and glasses generated 
with different cooling rates is important.
 When the  hyperquenched
 glass of water is properly annealed  at $T=130$ K\cite{naturejohari}, a
 reproducible weak endothermic transition is observed which has been
 associated with
 the calorimetric glass transition temperature $T_g$.
 For water, the standard cooling rate is not
 sufficiently fast to overcome crystallization, so active debate
\cite{angel1,johari,angel2} concerns how to relate the $T_g$ of
hyperquenched water to the unmeasurable $T_g$ of the slowly cooled glass.

In this work we use molecular dynamics (MD) simulations to address
the relation between liquid and glass configurations.
We reproduce the same procedure followed 
experimentally to generate glasses. We use two cooling rates $q_c$
 differing by almost three orders of magnitude.
 We also study the
heating rate and aging effects in the glass, an aspect novel to numerical
simulations.
 We work in the framework of the potential energy landscape (PEL) approach, in which
the $6N$-dimensional configurational space ---
defined by the $3N$ center of mass coordinates and by the $3N$ Euler  
angles  --- is  partitioned into a set of basins, each associated
with a different local minimum of the PEL \cite{sw,debenedettiNewNat}. 
We focus in particular on the depth of the local minimum closest to the system point, and on the
local curvatures of the PEL around the local minimum.  We discover that with
slow cooling rates, 
the glass retains a configuration very similar to a configuration sampled by
the liquid at higher $T$, and hence all structural properties of the glass can
be related to the structural properties of the liquid.
 In this case, a fictive temperature can be defined, the temperature at which the glass configuration is sampled by the liquid.
In the case of a fast cooling rate, aging phenomena are very 
active already during the cooling process. Dynamics moves the
configuration in regions of the PEL which are never explored in
  equilibrium \cite{mossa2}.
Further aging at low temperature increases the differences between the glass
and the liquid. 
When this is the case, the glass does not possess a structure corresponding to the equilibrium liquid at any temperature and hence it is not possible to associate an unique fictive $T$ with the glass configuration.


We perform MD simulations for a 
system of $N=216$  molecules at
 fixed density, $\rho =1$~g/cm$^3$, interacting with the SPC/E
potential\cite{berensen}, with periodic boundary conditions. Interactions are
cut off at a distance of $r=2.5 \sigma$ ($\sigma$ parameterizes the
Lennard-Jones part of the SPC/E potential) and reaction field corrections are added to account for the long range interactions. 
Quantities are averaged over 32 independent trajectories.   We perform three types of MD calculations:
(i) cooling scans at constant rate, starting 
 from equilibrium liquid configurations at $T=300 K$, 
(ii) heating scans at constant rate (from glass configurations at $\approx 5$
K), and (iii) aging runs at constant aging temperature
 (at $T_{\rm age}=100$~K and $T_{\rm age}=180$~K, where 
significant aging effects are observed). The two cooling rates are 
$q_c=-3 \times 10^{10}$~K/s and $q_c=-10^{13}$~K/s, and the
 two heating rates are $q_h=+3 \times 10^{10}$ K/s and $q_h=+10^{13}$ K/s.
We will denote by fast-quenched glass the glass obtained with the fast
 cooling rate
 and by slow-quench glass the glass obtained with the slow cooling
rate. An averaged slow scan requires a simulation lasting 320 ns, 
close to the maximum possible by our method.
This limitation prevent us from studying larger systems~\cite{footnote}.
The location of the system on the PEL is studied 
performing numerical minimizations (conjugate gradient algorithm)  along the runs to estimate the closest local minimum configuration or inherent structure ($IS$), its 
 energy  $e_{IS}$ and the set of  $6N$ eigenvalues ${\omega_i^2}$ of the Hessian matrix\cite{inm}.
The  local curvatures around the minimum define in the harmonic approximation
the multi-dimensional parabolic shape of each PEL basin.
As a global indicator of shape we use the shape function defined as
\begin{equation}
{\cal S}_{IS} \equiv \frac{\sum_{i=1}^{6N-3} ln(\hbar \omega_i/A_0)}{N}
\end{equation}
where $A_0=1$~kJ/mol.
At density $\rho =1$g/cm$^3$ the temperature dependence of  $e_{IS}$ and ${\cal S}_{IS}$ for the SPC/E model has been previously calculated\cite{newPRL}.
 
Focusing on landscape properties has the advantage that vibrational features are suppressed, and hence the remaining temperature dependence reflects 
structural changes.  The landscape approach is particularly useful for comparing properties of different configurations. 
Indeed, at constant volume, in equilibrium, all landscape
properties (i.e., $e_{IS}$, ${\cal S}_{IS}$) are functions only of
$T$. By eliminating $T$, we can establish a direct relation between 
${\cal S}_{IS}$ and $e_{IS}$,
i.e. between properties of the landscape sampled in equilibrium. 
A set of configurations with the same $e_{IS}$ can be considered
representative of equilibrium configurations if and only if ${\cal S}_{IS}$
satisfies the equilibrium ${\cal S}_{IS}(e_{IS})$ landscape relation.
 We compare the relation between shape and depth 
during cooling scans, heating scans and aging with the equilibrium relation. This 
allows us to test if  the region of configuration space explored during the scans is identical to the region explored by the equilibrium liquid at some higher temperature.

\begin{figure}[t]
\vspace{.2in}
\centerline {
\includegraphics[width=2.8in]{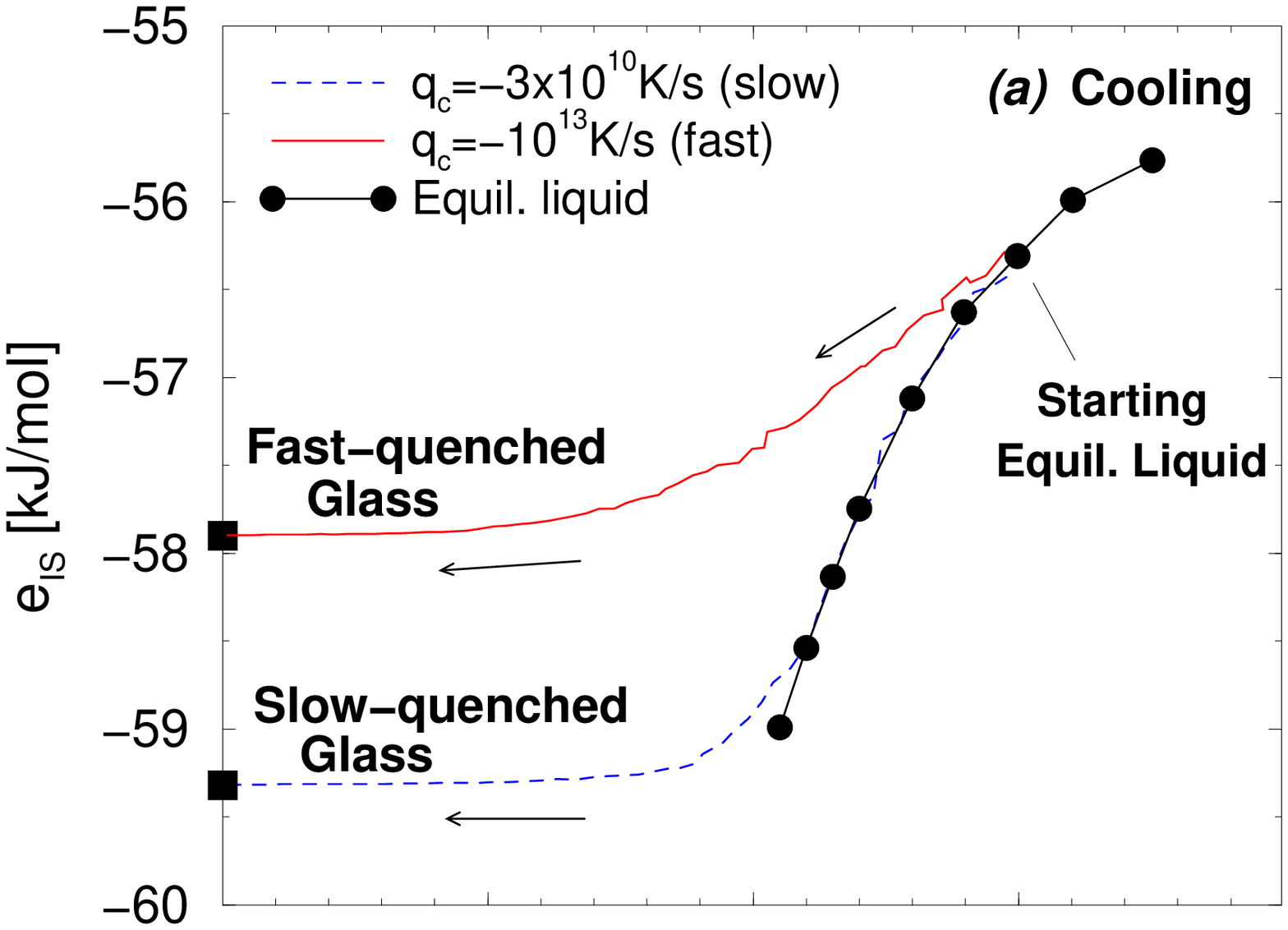}
}
\centerline {
\includegraphics[width=2.8in]{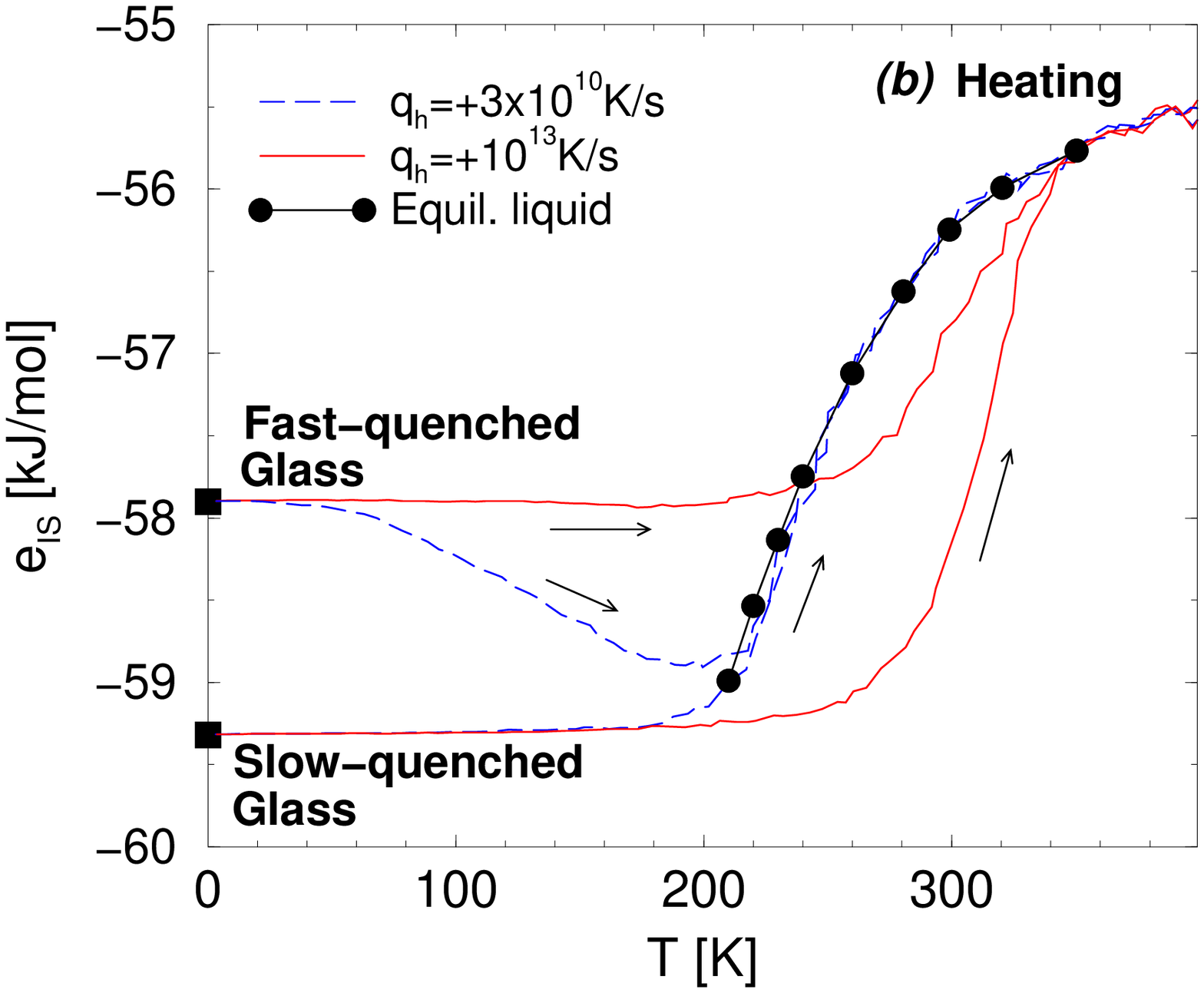}
}
\centerline {
\includegraphics[width=2.8in]{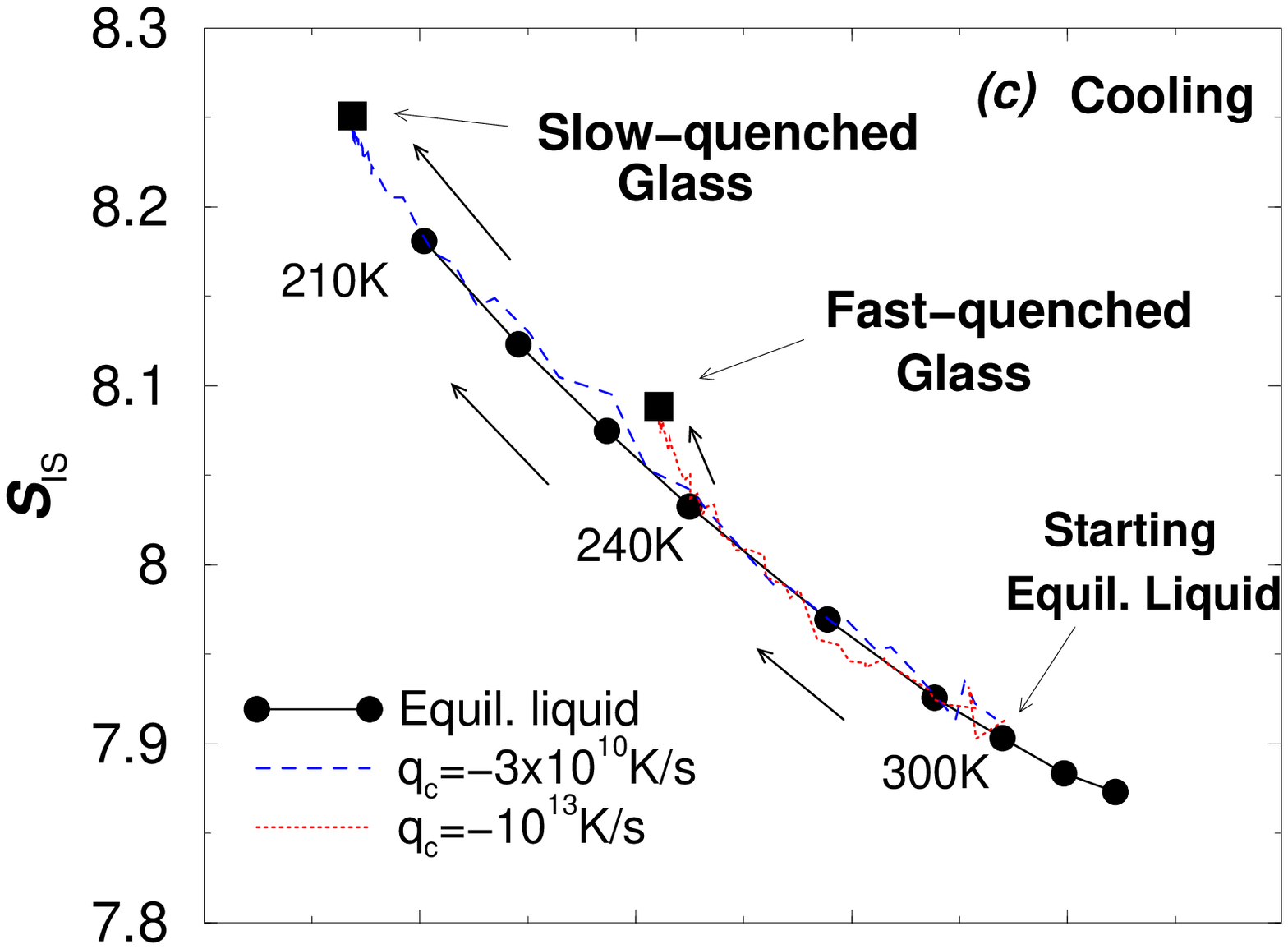}
}
\centerline {
\includegraphics[width=2.8in]{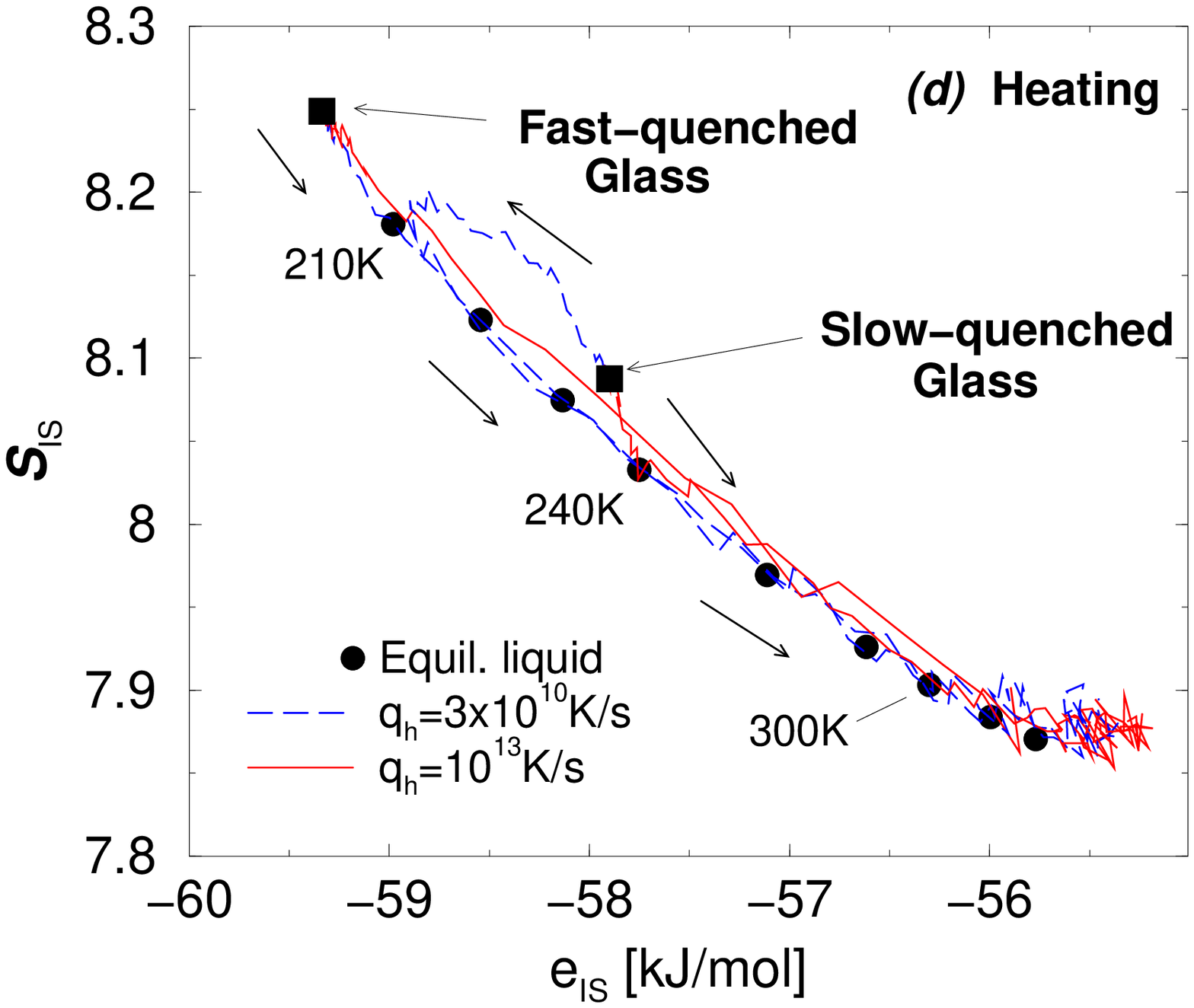}
}
\caption{(a) Cooling of equilibrium liquid
  configurations at $T=300$~K to $T \approx 0$~K with a fast
 ($q_c=-10^{13}$~K/s) and slow ($q_c=-3 \times 10^{10}$~K/s) cooling
 rate to generate fast-quenched and slow-quenched glasses, respectively.
 While during fast cooling the system is out of equilibrium already at
  $T=290$~K, slow cooling allows the system to sample the
  equilibrium IS as low as $T=220$~K.
(b) Heating of the fast-quenched and slow-quenched glasses with a fast
 ($q_h=10^{13}$~K/s) and slow ($q_h=3 \times 10^{10}$~K/s) heating rate.
 In the slow
 heating scan, the fast-quenched glass shows relaxation to deeper IS,
 a feature missing in the  
 slow-quenched glass case.  Fast heating shifts plots to higher $T$ and avoids
 relaxation in the fast-quenched glass.
(c) Basin shape function ${\cal S}_{IS}$  during fast and slow cooling. 
  For the slow cooling rate the system is able to follow the
 equilibrium ${\cal S}_{IS}(e_{IS})$, but for the fast cooling rate the system
 ends up outside the equilibrium curve.
(d) ${\cal S}_{IS}(e_{IS})$ during heating of the
  fast-quenched and slow-quenched glasses. During heating,
 the slow-quenched glass follows the ${\cal S}_{IS}(e_{IS})$ path for
 equilibrium liquids. In the case of the fast-quenched glass the system
 approaches the equilibrium ${\cal S}_{IS}(e_{IS})$ only at large times.}
\label{CoolingRates}
\end{figure}

Figure~\ref{CoolingRates}a shows $e_{IS}(T)$
 for the fast and the slow cooling rates, together with
equilibrium values from Ref.\cite{newPRL}.  The equilibrium data show that in the liquid state, the system explores deeper and deeper basins on lowering $T$ and that at each $T$, basins with a well-defined energy depth are populated. The fact that different depths are associated with different $T$ provides a mechanism for associating a configuration (whose $e_{IS}$ value is known) to a  fictive temperature $T_f$, as suggested in Ref.\cite{kob}.
In  the fast cooling case the liquid loses equilibrium at a very high
 temperature (around $290 $K), i.e., when the system is still sampling basins
 of high energy. In the slowest cooling case, the system loses
 equilibrium around $220$ K. 
 For the fast cooling case, the range of $T$ in which $e_{IS}$ changes is
 significantly larger, and more energy (from the $T$ where equilibrium is
 lost to the final $T$)
 is released during the cooling scan.

Figure~\ref{CoolingRates}b shows $e_{IS}$ for the fast and slow 
 heating rates starting both from the fast-quenched
glass and the slow-quenched glass. For the fast heating rate, 
to reach equilibrium $T$ must reach a value larger than the $T$ at which
 the same system lost equilibrium on cooling (Fig.~\ref{CoolingRates}a). In
 the slow heating case, a new phenomenon appears for the 
fast-quenched glass, the system starts to explore PEL basins
with lower and lower $e_{IS}$ as it relaxes toward equilibrium, significantly changing the location of the system in configuration space.

Figures~\ref{CoolingRates}c  and~\ref{CoolingRates}d show
 the ${\cal S}_{IS}(e_{IS}$) relation for the 
scans reported in Figs.~\ref{CoolingRates}a  and~\ref{CoolingRates}b, respectively.
  One sees that in the slow cooling case, the cooling scan proceeds via a sequence of equilibrium configurations 
(since  the ${\cal S}_{IS}(e_{IS})$ relation coincides with the equilibrium
 one), i.e. basins with same depth explored during the slow scan and those in
 equilibrium have also the same shape.
However, fast cooling generates a glass (the black square) which 
we see lies outside the equilibrium curve of Fig.~\ref{CoolingRates}c.
In this case, basins explored during fast cooling that have the same depth 
as those found in equilibrium have smaller curvatures.
 During the slow heating process
 ${\cal S}_{IS}(e_{IS})$ moves even further from the equilibrium line 
(Fig.~\ref{CoolingRates}d).

\begin{figure}[htb]
\vspace{.2in}
\centerline {
\includegraphics[width=2.8in]{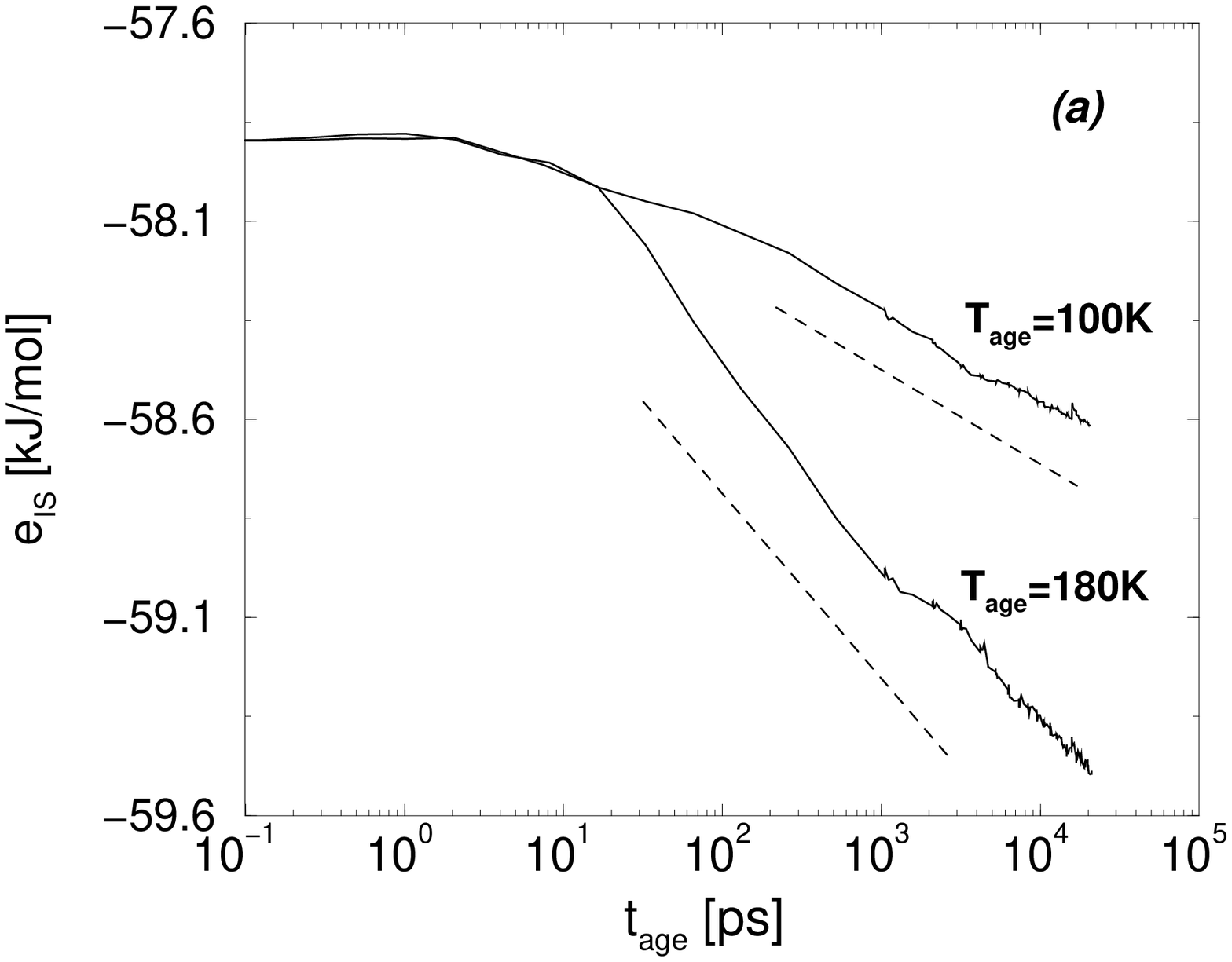}
}
\vspace{.2in}
\vspace{.2in}
\centerline {
\includegraphics[width=2.8in]{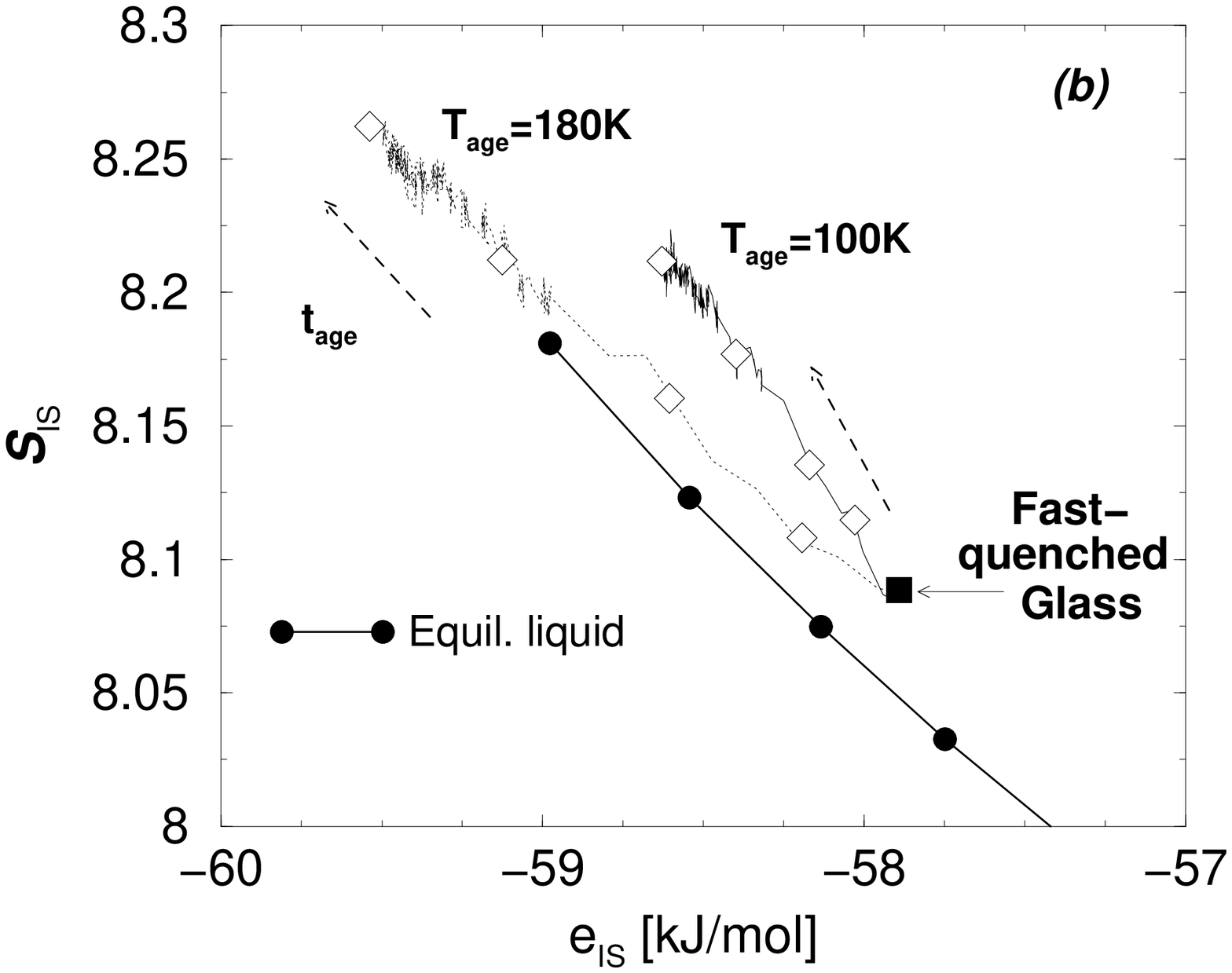}
}
\vspace{.2in}

\caption{(a) Time dependence of $e_{IS}$ and (b) relation between basin shape
  ${\cal S}_{IS}$ and basin depth during aging of the fast-quenched glass at
  two temperatures  $T_{\rm age}=100$~K and $T_{\rm age}=180$~K. For both
  cases, for
  long $t_{\rm age}$, $e_{IS} \approx \log(t_{\rm age})$ (dashed lines). Note that during aging at
  low $T_{\rm age}$, the system explores basins for which the relation between
  shape and depth does not correspond to the equilibrium
 ${\cal S}_{IS}(e_{IS})$ relation.  At large $T_{\rm age}$, aging
 moves the system closer to equilibrium. At each $T_{\rm age}$, open diamonds  correspond  to (from right to left)
$t_{\rm age}=20$~ps, $t_{\rm age}=200$~ps, $t_{\rm age}=2$~ns, and $t_{\rm age}=20$~ns.
}
\label{eis-tage}
\end{figure}

We now focus on the trajectory of the fast-quenched glass during aging at 
two different aging temperatures $T_{\rm age}=100 K$  and $T_{\rm age}=180 K$.
 For both rates we find that
$e_{IS}$ is a monotonically decreasing function of aging time $t_{\rm age}$
(Fig.~\ref{eis-tage}a),
 suggesting that the system changes basin by finding specific directions on
 the PEL which involve low energy barriers.
 For long $t_{\rm age}$, $e_{IS} \approx \log(t_{\rm age})$, as found in atomic liquids\cite{kob,aging}.
During aging, two different behaviors are again observed for the
PEL ${\cal S}_{IS}(e_{IS})$ relation (Fig.~\ref{eis-tage}b).
 In the high $T_{\rm age}$ scan, the
system approaches equilibrium configurations, while for low $T_{\rm age}$ the
opposite is seen. Note (Fig.~\ref{eis-tage}b) that aging at
low $T_{\rm age}$ moves the system even further from the equilibrium 
${\cal S}_{IS}(e_{IS})$ curve.
  
\begin{figure}[htb]
\centerline {
\includegraphics[width=2.8in]{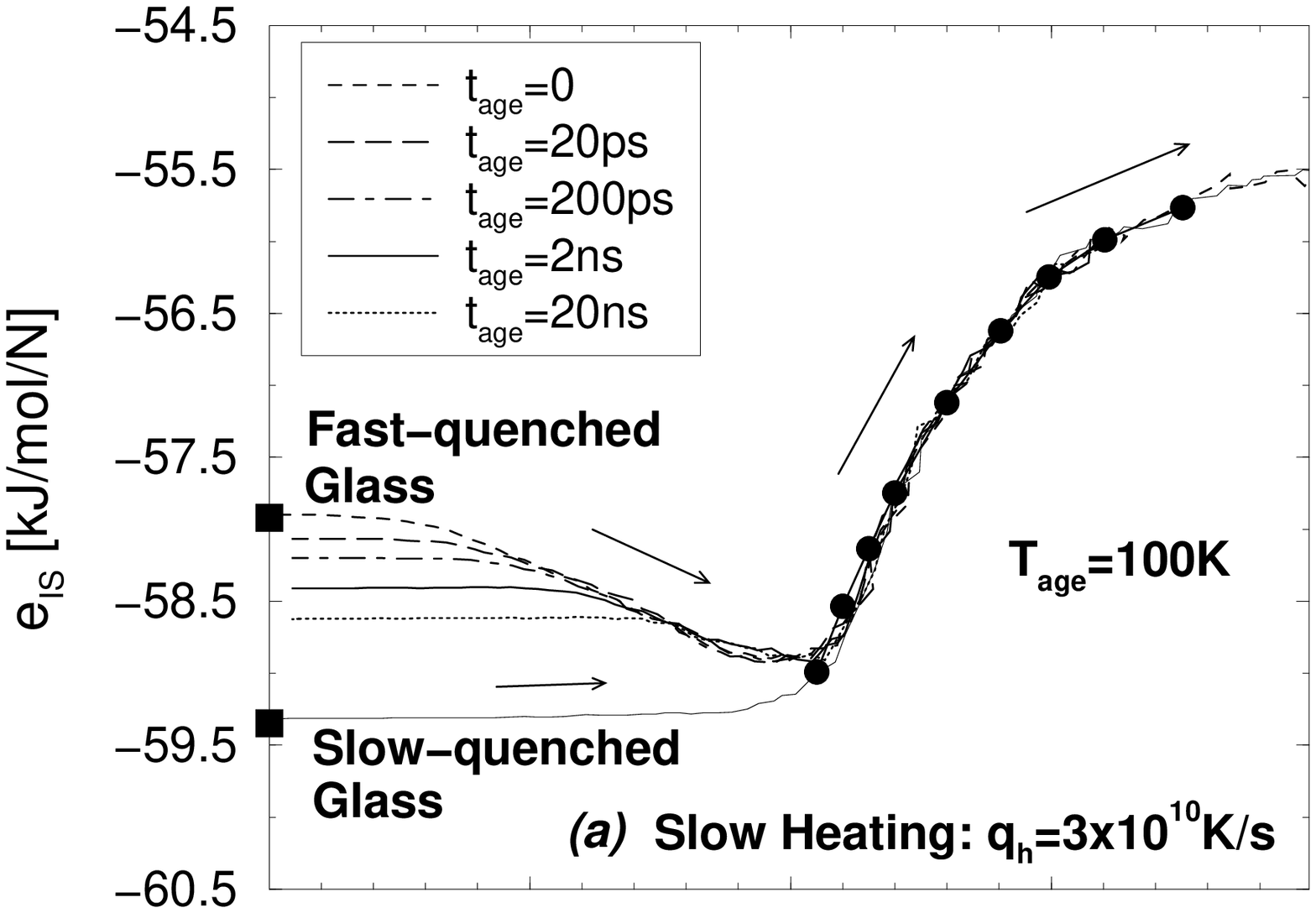}
}
\centerline {
\includegraphics[width=2.8in]{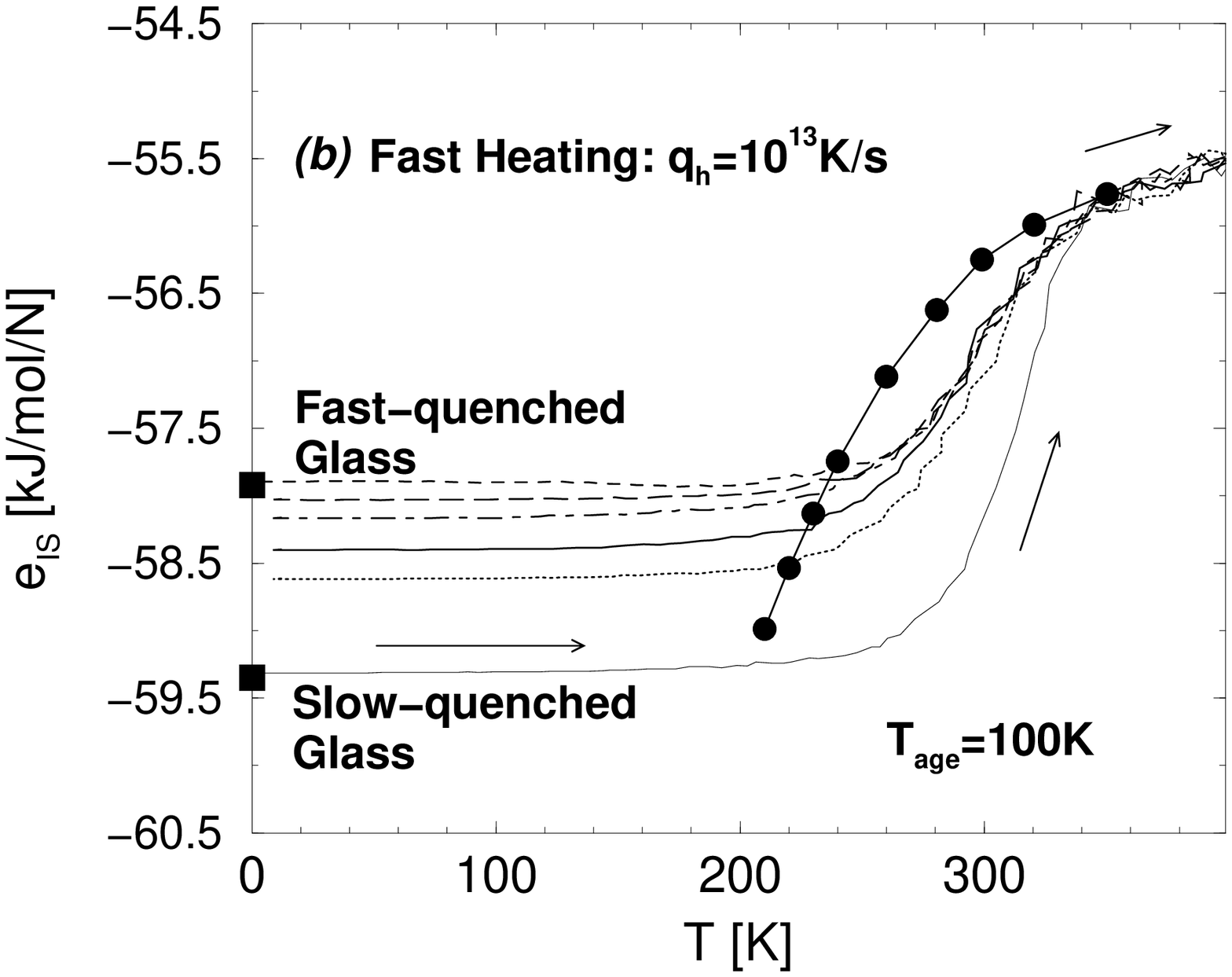}
}
\centerline {
\includegraphics[width=2.8in]{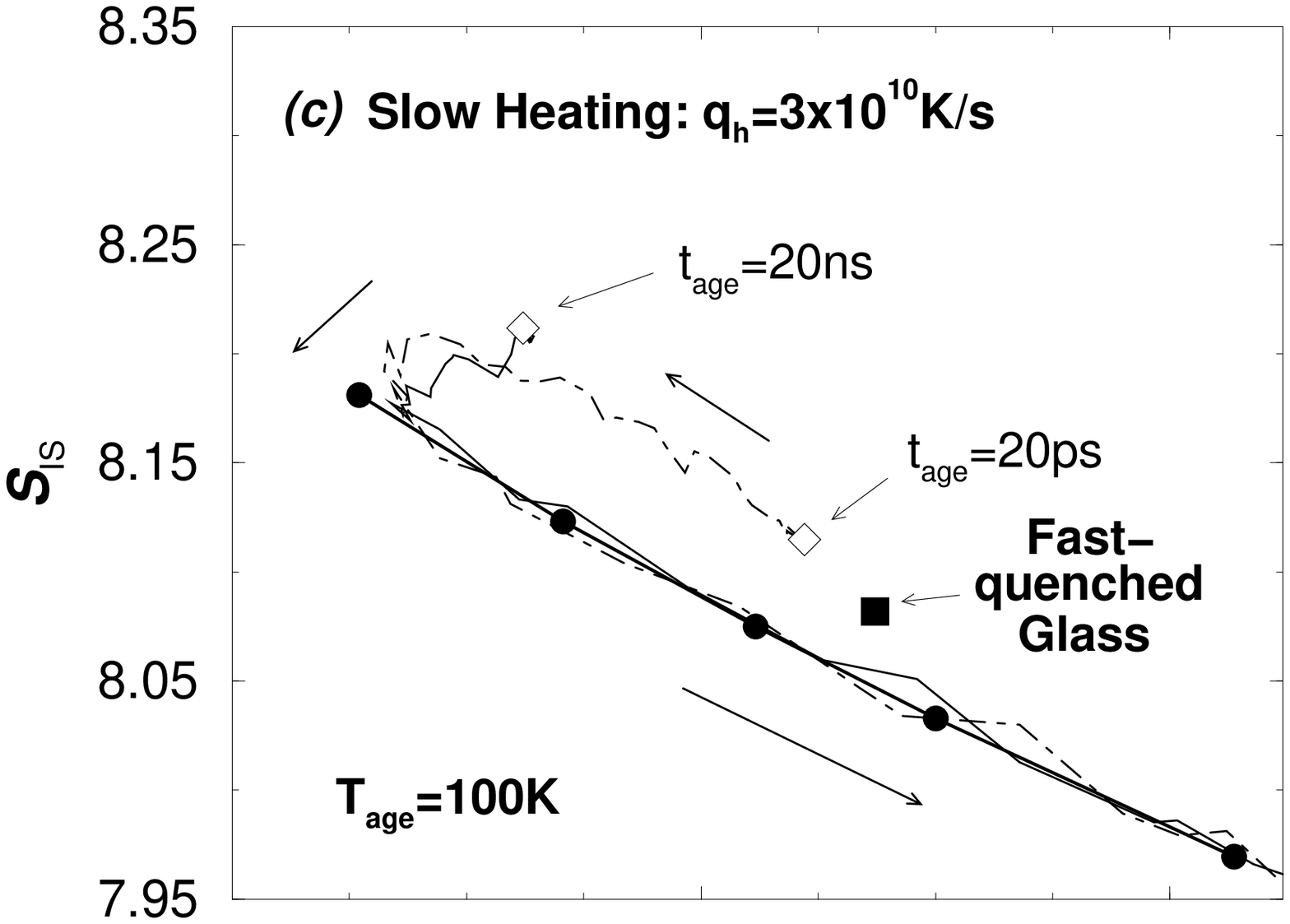}
}
\centerline {
\includegraphics[width=2.8in]{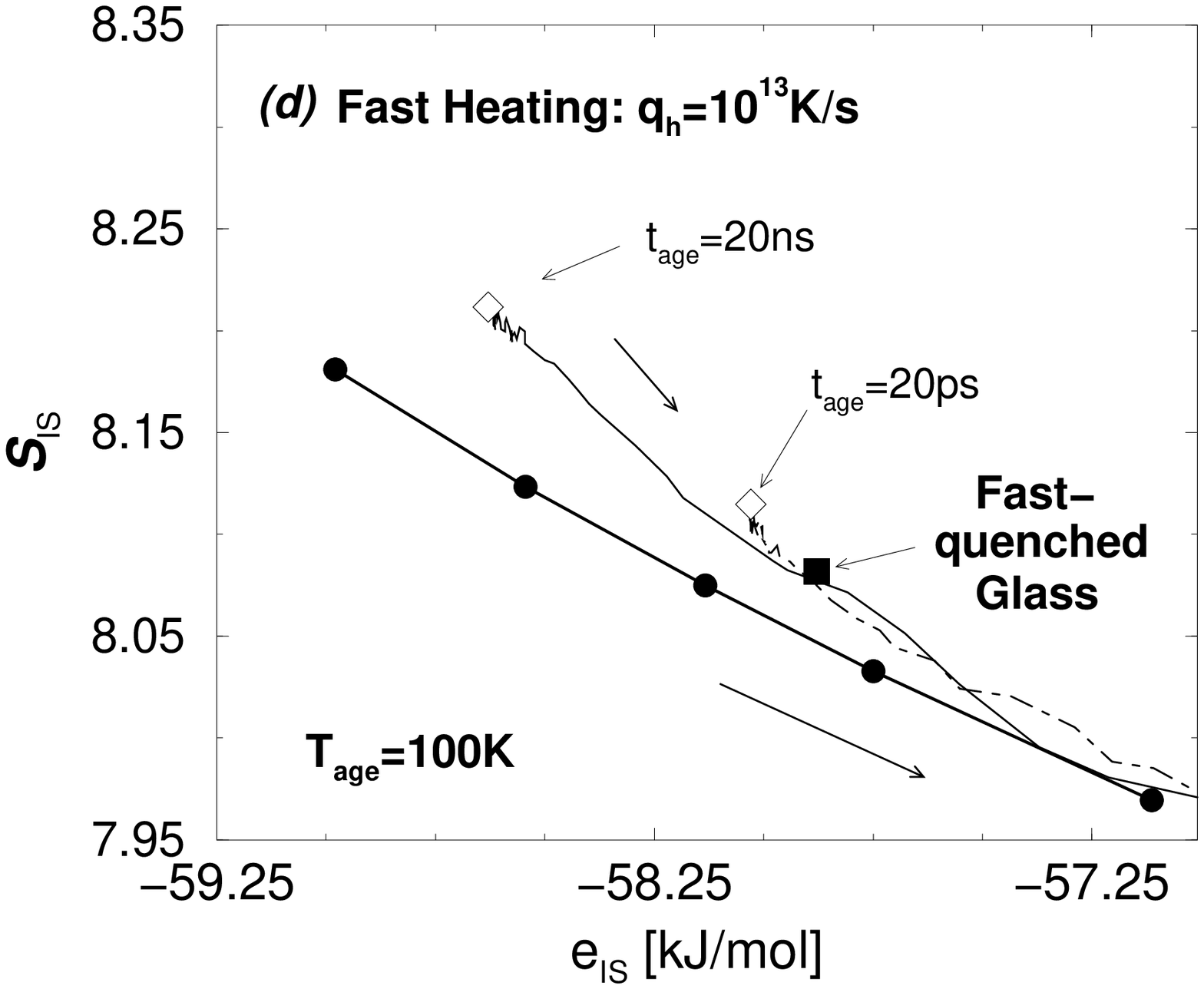}
}
\caption{Inherent structure energy $e_{IS}$ and basin shape function 
${\cal S}_{IS}$ during heating of the fast-quenched glass with (a,c) slow
and (b,d) fast 
heating rates after aging for  different times $t_{\rm age}$ at a temperature
$T_{\rm age}=100$~K. Results for T=180 K are qualitatively similar.
Filled circles correspond to data for equilibrium liquid
configurations from Ref.~[19]
 For comparison, in (a) and (b) we show $e_{IS}$ during the
heating scan from the slow-quenched glass. In (c) and (d), we only show 
${\cal S}_{IS}$ in the heating scans after aging for $t_{\rm age}=20$~ps
 and $t_{\rm age}=20$~ns. }
\label{T100}
\end{figure}

Next we show (Fig.~\ref{T100}) heating scans for different aging times 
after isothermal aging at $T_{\rm age}=100$~K.  Data for $T_{\rm age}=180$~K
(not shown) are qualitatively similar. The IS energy of the
starting configuration is lower the longer the time spent at
$T_{\rm age}=100$ K before beginning to heat.
 We see (Fig.~\ref{T100}c) that $e_{IS}$ is monotonic on heating.
Figure~\ref{T100}b shows the results for the fast heating
rate. In this case, $T$ increases so fast that the system does
not have time to find deeper IS. Therefore, there are no aging effects
and $e_{IS}$ grows monotonically with $T$.  From Figs.~\ref{T100}a
and~\ref{T100}b,
 we see that for the fast heating rate, equilibrium is reached at  a much
 higher  temperature compared to the case of the slow heating rate, in full
 agreement with the no-aging ($t_{\rm age}=0$) case, as observed in Fig.~\ref{CoolingRates}b.  The aging effects on ${\cal S}_{IS}(e_{IS})$ are
shown in Figs.~\ref{T100}c (for slow) and~\ref{T100}d (for fast) heating
 rates, confirming that in the slow
 heating case the fast-quenched glass keeps exploring new basins
 never seen in equilibrium before meeting the equilibrium curve.

In summary, we have shown 
that the fast-quenched glass significantly differs from the slow-quenched
glass. To the extent that this model captures key features of real water, our
work provides  support to the idea that, if crystallization would not interrupt the cooling process in water, 
the slow-quenched glass would be characterized by a different $T_g$ compared to
the fast-quenched glass~\cite{angel1,johari,angel2}. The analysis of
simulated configurations  has allowed us to clarify the differences between
the slow- and the fast-quenched glass.  The differences lie not only in the expected difference in $e_{IS}$ but,  more significantly, in the fact that the fast-quenched glass during cooling loses contact with the liquid state and starts to explore regions of the PEL which are never explored in equilibrium.  
The fast-quenched glass does not possess a structure corresponding to the equilibrium liquid structure at any temperature, and hence it is not possible to associate with it a unique fictive temperature. Finally, we have found that aging
in the glass phase  provides a mechanism for reducing  differences 
between the liquid and the glass structures only if the temperature at which the system ages is sufficiently high. Remarkably, at low $T_{\rm age}$,  differences between the liquid and the glass structures appear to increase with $t_{\rm age}$.

We thank C.A. Angell and S. Sastry for discussions, the BU Computation Center
 for CPU time, and the NSF Chemistry Program, MIUR Cofin 2002, Firb, and
 INFM Pra GenFdt for support.

\begin{references}
\bibitem{inm}
The Hessian matrix is the matrix formed by the second derivatives of the
 potential energy with respect to the center of mass coordinates and
 Euler angles of each molecule.
\end{references}
\end{document}